# Magneto-Optical Resonance of Electromagnetically Induced Absorption with high contrast and narrow width in a vapour cell with buffer gas

D.V. Brazhnikov[1,2], A.V. Taichenachev[1-3] and V.I. Yudin[1-4]

[1] Institute of Laser Physics SB RAS, pr-kt Lavrent'eva 13/3, Novosibirsk, 630090, Russia
[2] Novosibirsk State University, ul. Pirogova 2, Novosibirsk, 630090, Russia
[3] Russian Quantum Center, ul. Novaya 100A, Skolkovo, Moscow Region, 143025, Russia
[4] Novosibirsk State Technical University, pr-kt K. Marksa 20, Novosibirsk, 630073, Russia

**Abstract.** The method for observing the high-contrast and narrow-width resonances of electromagnetically induced absorption (EIA) in the Hanle configuration under counterpropagating light waves is proposed. We theoretically analyze the absorption of a probe light wave in presence of counterpropagating one with the same frequency as the function of a static magnetic field applied along the vectors of light waves, propagating in a vapour cell. Here, as an example, we study a "dark" type of atomic dipole transition $F_g=1 \rightarrow F_e=1$ in $D_1$ line of $^{87}$Rb, where usually the electromagnetically induced transparency (EIT) can be observed. To obtain the EIA signal one should proper chose the polarizations of light waves and intensities. In contrast of regular schemes for observing EIA signals (in a single travelling light wave in the Hanle configuration or in a bichromatic light field consisted of two traveling waves), the proposed scheme allows one to use buffer gas to significantly enhance properties of the resonance. Also the dramatic influence of atomic transition openness on contrast of the resonance is revealed, that gives great advantage in comparison with cyclic atomic transitions. The obtained results can be interesting in high-resolution spectroscopy, nonlinear and magneto-optics.

## 1. Introduction

At first time phenomenon of coherent population trapping (CPT) was discovered in the experiment [1]. Since that time there were a lot of theoretical and experimental works devoted to various aspects of the phenomenon (see, for example, [2-4]). The sense of CPT is that a resonant laser radiation, interacting with an atom, can drive it into a special coherent state. Atoms, being in this state, do not absorb and scatter the photons from driving resonant light field any more. As a result the fluorescence of atomic gas tends to be zero. That is why this coherent state is often called "dark state". To date CPT has been found many interesting applications in laser physics [5], precision laser spectroscopy [6,7], quantum informatics (dark-state polaritons and quantum memory) [8], laser cooling below the recoil limit [6,9] and the others. CPT is accompanies by steep dispersion of refractive index, what has been found applications in nonlinear optics [10] and optical communications [11-13]. Especially one more application should be noted – it is a quantum metrology (producing of miniature atomic clocks and magnetometers with high sensitivity and noise immunity, low energy consumption [14-17]). This direction promises great prospects for various mobile navigational, communicational, data transferring and measurement systems. The general principle of such hi-tech devices concerns with the resonance of electromagnetically induced transparency (EIT) [18], that is nothing but the spectroscopic manifestation of CPT phenomenon. In fluorescence signal from a vapour cell the EIT resonance is usually observed as a narrow dip. So that EIT signal is also called "dark resonance". One of the main brilliant and high demanded features of EIT resonance consists in its width, which can be much smaller than the natural linewidth. Resonances owning this feature are often called "subnatural-width" resonances. Modern spectroscopic methods allow one to obtain quite narrow width of the subnatural-width resonance down to a few Hz [7,17,19]. Perhaps, these resonances are the narrowest ones that can be observed in a cell filled with "hot" atomic vapour (i.e. at room temperature).

In 1997 subnatural-width resonance with opposite sign was discovered – electromagnetically induced absorption (EIA, [20,21]). Firstly that resonance was observed under a bichromatic laser field composed of co-directional beams with opposite circular polarizations. Then the effect was also studied under a single-frequency light wave accompanied with a static magnetic field applied along the wave vector (magneto-optical or, so-called, Hanle configuration) [22]. Since its discovery that type of subnatural-width resonances ("bright" resonance) has also been found some applications ("fast" light [23], four-wave mixing [24], magnetic field mapping of cold atomic samples [25]). However, scope of EIA applications happened to be rather small due to some difficulties. The fact is that the most effective methods for getting the resonance width narrower based on the usage of buffer gas or a cell with antirelaxation coating of walls. These methods, demonstrating perfect results for EIT signals [19,26], happened to be useless for EIA ones in the standard schemes of observation [20-22]. It is due to the fact that EIA in those schemes is resulted from formation of anisotropy on an atomic exited state and its spontaneous transfer to ground one [27-29]. Collisional process between atoms and buffer gas or cell walls rapidly destroys the exited state anisotropy, what leads to damping of EIA signal or even changing its sign (EIA→EIT) [30,31]. For a long time the best results for the resonance width were about 10 kHz simultaneously with the low contrast of a few percents (see, for example, [21,22]). Those results rather yield to the results with EIT signals, which could have simultaneously the contrast of 40-90% and sub-kHz widths (e.g. [32-34]).

In the series of papers [35-37] we have proposed some "unconventional" scheme for observing EIA signals, namely we have studied the effective polarization method for controlling a sign of subnatural-width resonance (EIT↔EIA) in the Hanle configuration under counterpropagating light waves. Then we figured out that the method can be effectively exploited for observing magneto-optical EIA resonances with rather good properties (contrast and width), no worse than EIT signals and even better. It worth noting here that there are the other "unconventional" schemes for observing EIA signals with enhanced properties, but many of them is comparatively complicated. In particular, authors of Ref. [38] proposed to observe EIA signals with the help of microwave field applied between ground sublevels of Λ-scheme. Also EIA resonances can be observed using Ramsey technique with complex light-beam cross-section profile: in paper [39] the pump beam had a hole, in which the probe wave was injected. In the experiments of Ref. [40] the bichromatic pump laser field composed of counterpropagating waves was used. The contrast of EIA signal in probe-wave transmission no more than 3% was detected at sufficiently large width about 250 kHz. The authors of experimental work [41] observed the resonance in the Hanle configuration using "bright" atomic transition $F_g=2 \rightarrow F_e=3$ on $D_2$ line of $^{87}$Rb, placed in a cell with antirelaxation coating. To observe EIA resonance in the probe wave absorption they were in need of additional transverse magnetic field. In the last work the authors were able to achieve sufficiently narrow EIA signal (about 390 Hz), but simultaneously with low contrast (units of percents). There are the other detailed experimental and theoretical studies of using transverse magnetic field to acquire EIA resonances on "dark" transitions (see, for instance, [42]). In our work [43] we suggested to use a traveling light wave with optimal ellipticity of polarization in regular Hanle scheme. It helped to increase EIA resonance amplitude in times, but the contrast was also too small at rather big width (of the order 1 MHz). The different shapes of laser beam profile also do not improve the situation with contrast [44]. Some improvement of EIA contrast can be achieved by using atomic beams instead of vapour cells. For instance, authors of Ref. [45] used collimated $^6$Li beam under bichromatic laser field, driving $D_2$ line closed atomic transition (i.e. $F_g=3/2 \rightarrow F_e=5/2$). The contrast of EIA resonance in the fluorescence signal as much as 40% was observed, but the width was also too large (hundreds of kHz). So, in spite of really large variety of modern schemes for observing EIA resonances, we have to conclude that none of them are able provide high contrast (50% and more) accompanied with reasonably narrow width (kHz and sub-kHz levels).

In this paper we describe the idea of a new observing scheme on the example of an open dipole transition $F_g=1 \rightarrow F_e=1$ (it can be implemented, for instance, on $D_1$ line of $^{87}$Rb). Here $F_g$ and $F_e$ are the total angular momenta of the atom for ground and exited states respectively. The proposed scheme allows one to use buffer gas (or a cell with antirelaxation coating of walls) to greatly enhance the resonance properties. In our case buffer gas does not lead to damping of EIA as in regular schemes. In the proposed scheme the EIA signal results from the interaction of a probe beam with a ground atomic state that has been specially prepared by the pump (counter-directed) laser beam. Anisotropy of the exited state is destroyed rapidly due to buffer-gas-caused collisional depolarization. Its absence does not play crucial role for observing EIA signal in contrast of the regular schemes of observation [30,31]. Besides, openness of the transition can result in EIA resonance with very high contrast up to 100%. At that the spectroscopic profile consists of almost only subnatural-width spike without any wide resonance structure (saturated-absorption, Doppler or other contributions into the profile). In other words, in a probe absorption signal the ratio between EIA amplitude and background signal value becomes dramatically large (up to $10^4$ and more percents). It can lead, for optimal pump wave intensity and atomic concentration in a cell, to very narrow and high-contrast EIA resonance in a transmission signal after the cell. In this consists the main difference between open and closed (cyclic) atomic transitions, where the contrast no more than 10% can be achieved. Besides, it is very difficult to reach the similar effect of 100%-contrast for EIT resonances on open atomic transitions, because, as a rule, the openness makes EIT contrast much worse [46].

We also would like to note that one is the main features of the method proposed here is its comparative simplicity: there is no need in microwave radiation or bichromatic light fields, additional transverse magnetic field or complex geometry of light field. We also suggest using just a vapour cell that is much simpler than using an atomic beam as in Ref.[45]. Besides, the new method gives an opportunity, if necessary, for effective and easy control of the resonance sign (EIT↔EIA), simply by changing the angle between linear polarizations of two counterpropagating laser beams (pump and probe ones). Theoretical calculations demonstrate the principal achievement of EIA resonances in transmission signal, having width on the level of 300 Hz and contrast very close to 100%.

## 2. Bright resonance on the dark transition

In the papers [35-37] we have experimentally and theoretically studied the new method for observing EIA signals on "dark" transitions, where coherent population trapping can be induced by the resonant laser radiation ($F_g=F \rightarrow F_e=F$, with F an integer, and $F_g=F \rightarrow F_e=F-1$, with F an integer or a half-integer). In the cases of conventional observing schemes (e.g. [20-22,47]) only the resonances of EIT type can be acquired on these transitions. The physical origins of the sign transformation effect (EIT↔EIA) in the proposed scheme have been analyzed in details in the papers [35-37], where the qualitative explanation of sign-transformation effect has been demonstrated on the basis of regular Λ-model of an atom. Therefore, we omit the description here. Let us just note that, perhaps, the main value of the method is not so much the observing of the transformation EIT↔EIA, but in obtaining the high-quality EIA resonances. In the present work we demonstrate power of the method for the particular case of the open dipole transition $F_g=1 \rightarrow F_e=1$ (for instance, on $D_1$ line of $^{87}$Rb, see figure 1).

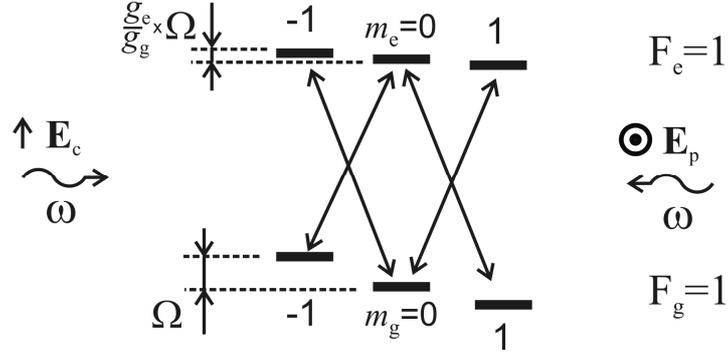

**Figure 1.** The dipole transition $F_g=1 \rightarrow F_e=1$ under the counterpropagating laser beams with orthogonal linear polarizations. The vertical solid lines denote $\sigma^+$ and $\sigma^-$ transitions induced by the waves.

So, let us consider the atom placed in a longitudinal static magnetic field and two light waves with the same frequency, propagating along the *z*-axis in opposite directions:

$$\mathbf{E}(z,t) = E_c \mathbf{e}_c \, e^{-i(\omega t - kz)} + E_p \mathbf{e}_p \, e^{-i(\omega t + kz)} + \text{c.c.} \;, \tag{1}$$

$$\mathbf{B} = B \mathbf{e}_z \;. \tag{2}$$

Here $E_c$ and $E_p$ are the scalar amplitudes of pump (c) and probe (p) waves, which are assumed to be real values without loss of generality, $\mathbf{e}_c$ and $\mathbf{e}_p$ are the complex unit vectors of pump and probe waves polarizations; "c.c." means complex conjugate term; $B$ is the amplitude of magnetic field applied along the Cartesian ort $\mathbf{e}_z$. The basis of our numerical calculations is the standard quantum-mechanical formalism of the density matrix, involving quantum theory of angular momentum of an atom (see, for example, monographs [48,49] and [37]). The Lindblad equation [50] on the atomic density matrix can be written in the form:

$$\left( \frac{\partial}{\partial t} + \upsilon \frac{\partial}{\partial z} \right) \hat{\rho} = -\frac{i}{\hbar} \left[ \left( \hat{H}_0 + \hat{H}_{EB} \right), \hat{\rho} \right] + \hat{\Re}\{\hat{\rho}\} \;, \tag{3}$$

with $\upsilon$ the *z*-projection of atomic velocity, $\hat{H}_0$ the Hamiltonian of a free atom, $\hat{H}_{EB}$ the operator of dipole interaction between an atom and the fields (1) and (2). Stochastic operator $\hat{\Re}\{\hat{\rho}\}$ is to take into account spontaneous relaxation, time-of-flight relaxation and various processes due to collisions.

The approximation of strong collisions [48] is involved in our theory to take into account buffer gas effects. Various processes, occurring due to collisions (exited state depolarization, dephasing of atomic polarization, velocity changing), and finite time of atom-field interaction are described by the corresponding relaxation constants. Such the way is widely used, because of its relative simplicity and clearness, and it qualitatively reflects main spectroscopic features of interaction between atoms, buffer gas and electromagnetic fields. The strict collisional theory is much more sophisticated and can be found, for instance, in monograph [48] (and references in it). In the present paper for short and ease of understanding we avoid the mathematical formalism of our method, but the numerical calculations are presented, showing the main idea and new effects.

As we have already mentioned above the particular open dipole transition $F_g=1 \rightarrow F_e=1$ is under consideration (figure 1). The transition properties are taken as for the real atomic transition $5^2S_{1/2}$, $F_g=1 \rightarrow 5^2P_{1/2}$, $F_e=1$ in $^{87}$Rb and are the follows: $\lambda \approx 795$ nm, radiative relaxation rate $\gamma \approx 2\pi \times 5.57$ MHz, branching coefficient $\beta=1/6$ defines the degree of transition closeness, Lande factors of energy levels $g_g = -1/2$, $g_e = -1/6$. Figure 1 includes the notation $\Omega = g_g \mu_B B/\hbar$ – Larmor frequency for the ground state (frequency of Zeeman splitting of the ground state among magnetic sublevels $m_g$), with $\mu_B$ the Bohr magneton. For the numerical calculations we have assumed some regular experimental conditions: buffer gas pressure ≈ 10–20 Torr, vapour temperature ≈ 300K, cross section of the light beams ≈ 5 mm. So, the relaxation constants may have the following estimates: rate of velocity changing collisions $\gamma_c \approx 50\gamma$ (we assume $\gamma_c$ to be real value and equal for ground and excited states of an atom, as well as for the optical coherences), rates of depolarization of the exited state and polarization dephasing are of the same order, i.e. $\gamma_e \sim \gamma_{deph} \sim \gamma_c$, rate of establishment of thermodynamic equilibrium among the ground state sublevels (inverse ground-state anisotropy lifetime) $\Gamma \sim 10^{-5}\gamma$. The last, of course, strongly depends on the certain buffer gas and pressure and it can vary in wide range. Some description of collisions by the relaxation constants may be found, for instance, in [51]. Also, in the limit $\gamma_c \gg \gamma \gg \Gamma$, what usually happens for spectroscopy with buffer gas, the used collisional model is similar to that described by E. Arimondo in [52]. Besides, since our theoretical model takes into account various relaxation processes, it is also appropriate for the case of a cell with antirelaxation coating of walls. That case demands separate study and we do not consider it here.

For better clarity of proposed observing configuration let us show a schematic representation of the main optical part of possible experimental setup (figure 2). Two travelling light waves with orthogonal linear polarizations are injected to a vapour cell from opposite windows. Some splitting plate (for instance, polarizing beam splitter) should be used to direct probe wave to a detector for analyzing after passing the cell. External stray magnetic field must by suppressed by multilayer shield or some active system based on Helmholtz coils. Spectroscopic signal (probe wave transmission) is monitored with scanning the solenoid voltage.

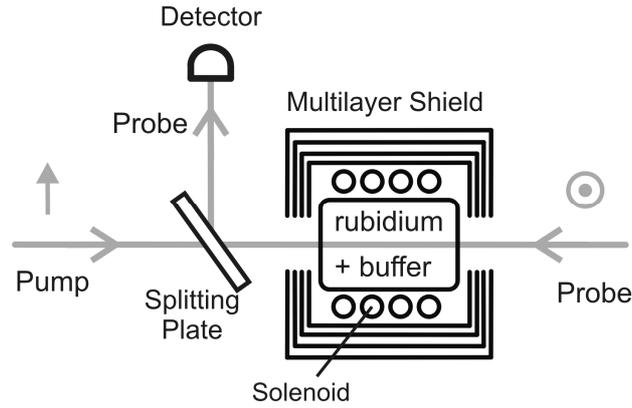

**Figure 2.** Main optical part of possible experimental setup.

We assume the probe wave to be much less than the pump one, i.e. $R_c \gg R_p$ with $R$ the Rabi frequency for corresponding wave. With this condition the absorption coefficient of the medium may be considered not depending on prove wave intensity. So, the probe power after the cell is

$$P(\Omega, R_c, z = L) = P_0 e^{-\alpha(\Omega, R_c)nL} \quad . \tag{4}$$

Here $L$ is the cell length, $P_0$ is the initial probe-wave power, $n$ is the atomic density (it can take values in wide range $\sim 10^8$–$10^{12}$ cm$^{-3}$ depending on cell temperature), and $\alpha$ is the absorption coefficient for unit concentration and length. The exponential law (4) is valid, if the absorption coefficient $\alpha$ does not depend on coordinate $z$. In our case such assumption is quite permissible, because pump light wave interact with open dipole transition in an atom and when condition (5) is satisfied the medium is almost transparent for the pump beam.

Since the atomic transition is open, the absorption coefficient strongly depends on the pump field even if it is rather small ($R_c \ll \gamma$). It is explained by long interaction time between work atoms and light field owing to buffer gas. In other words, even than $R_c \ll \gamma$ the number of scattered photons during atom flight is

$$\tau \gamma S \gg 1, \tag{5}$$

with $\tau \sim \Gamma^{-1}$ and $S$ the saturation parameter at the exact one-photon resonance condition: $S = R_c^2/\gamma_{eg}^2$, where $\gamma_{eg} = \gamma/2 + \gamma_c + \gamma_{daph} + \Gamma \gg \gamma$ is the relaxation rate of optical coherences (nondiagonal density matrix elements between exited and ground energy levels). The condition (5) also means that all nonstationary population processes among magnetic sublevels of ground state, which happen when an atom cross the light field, can be omitted.

On the basis of reduced Maxwell equation and (3) the absorption coefficient from (4) can take the form:

$$\alpha = \alpha_0 \left\langle \mathrm{Re}\left\{ \mathrm{Tr}\left[\left(\hat{V}\hat{\rho}^g - \hat{\rho}^e \hat{V}\right)\hat{N}\right]\right\}\right\rangle_\upsilon, \tag{6}$$

with $\alpha_0 = 3\lambda^2/4\pi$, $\lambda$ the probe wavelength, $\hat{V}$ the dimensionless operator:

$$\hat{V} = \sum_{q,m_g,m_e} e_p^q \, C^{F_e,m_e}_{F_g,m_g;1q} |F_e,m_e\rangle\langle F_g,m_g|. \tag{7}$$

Here $e_p^q$ is the $q$-component of the polarization vector $\mathbf{e}_p$ in circular basis (with $q = -1,0,+1$), $C^{F_e,m_e}_{F_g,m_g;1q}$ are the Clebsch-Gordan coefficients and $\langle\ldots|$, $|\ldots\rangle$ are the Dirac bra and ket vectors respectively. The dimensionless elements of matrix $\hat{N}$ are the following:

$$N_{kj} = \frac{\gamma V^*_{jk}}{\gamma_{eg} - i\left\{\delta_p + \Omega\left(F^g_{kk} - \dfrac{g_e}{g_g} F^e_{jj}\right)\right\}}, \tag{8}$$

where indexes $k$ and $j$ can take the values $1\ldots 2F_g+1$ and $1\ldots 2F_e+1$ respectively; $F^g_{kk}$ and $F^e_{kk}$ are diagonal elements of $z$-component of atomic angular momentum of ground and exited states (in $\hbar$ units). Notation "Tr[...]" in (6) means the trace operation, while angular brackets $\langle\ldots\rangle_\upsilon$ denote averaging over the Maxwellian velocity distribution.

Figure 3 presents the EIA resonance in normalized absorption coefficient, numerically calculated basing on equations (3) and (6). $A_1$ in figure 3 is the resonance amplitude and $A_2$ is the wide background level ($\Delta_{back} \approx 2\pi \times 220$ MHz for the case of figure 3). Let us introduce the ratio $C_\alpha = (A_1/A_2) \times 100\%$ – the contrast of EIA signal in respect to the wide background. As it is seen, the subnatural resonance can exhibit quite good contrast at narrow width: $\Delta_{EIA} \approx 2\pi \times 370$ Hz (FWHM) and $C_\alpha \approx 1000\%$.

As it is seen from figure 4, the probe-wave absorption coefficient $\alpha$ demonstrates excellent resonance features. Namely, at low intensity of pump wave ($R_c \sim 0.1\gamma$) EIA width can have the value of just several hundreds of Hz (inferior limit is $\approx \Gamma$). With increasing pump wave intensity, the contrast $C_\alpha$ is also increases up to very high levels $\sim 10^4$ % (ultra-high contrast). As the preliminary calculations have been shown, the effect of ultra-high contrast does not observe in case of cyclic transition (for instance, $5^2S_{1/2}$, $F_g=2 \to 5^2P_{3/2}$, $F_e=3$ on D$_2$ line in $^{87}$Rb that was investigated in [41] for a cell with coating of walls). The comparison study of EIA resonances in the proposed configuration for various types of atomic transitions will be carried out separately.

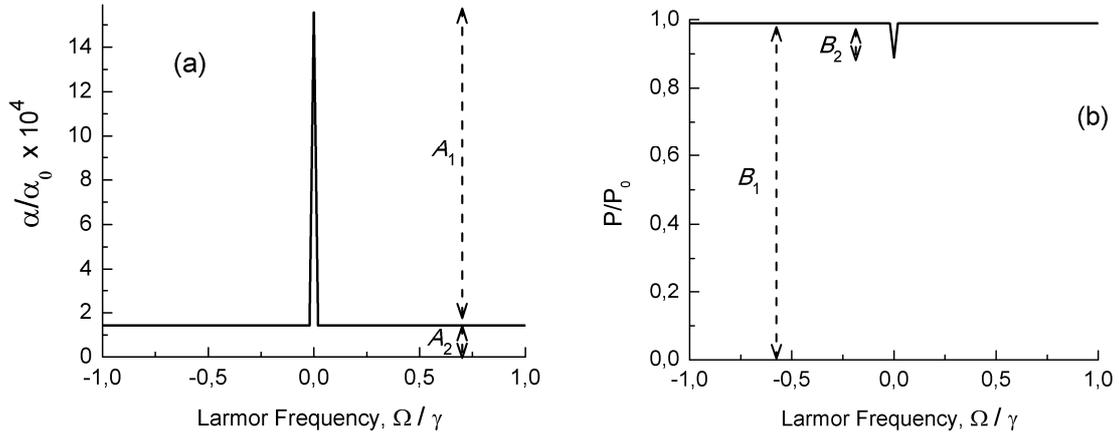

**Figure 3.** Ultra-high-contrast EIA resonance in the probe-wave absorption coefficient (a) for the open atomic transition $F_g=1 \to F_e=1$ on $D_1$ line in $^{87}$Rb. (b) Transmitted probe-beam power, corresponding to the case of figure 3a. Parameters are as follows: Rabi frequencies $R_c=0.2\gamma$ and $R_p=0.01\gamma$, relaxation rates $\gamma_c=50\gamma$ and $\Gamma=10^{-5}\gamma$, cell length $L=5$ cm and atomic density $n=10^{10}$ cm$^{-3}$.

The revealed effect of ultra-high contrast of EIA resonance in the probe-wave absorption coefficient is connected with optical pumping process. Indeed, in the absence of coherent population trapping in the ground state $F_g=1$ (far from $\Omega=0$, the profiles' wings) population of the exited state $F_e=1$ increases with the pump-field intensity increasing. Then, due to the spontaneous decay channel $F_e=1 \to F_g=2$, atoms populate the non-resonant ground level $F_g=2$, which starts to accumulate non-absorbing atoms. All that results in significant decrease of absorption of probe wave, which is resonant only to the transition $F_g=1 \to F_e=1$. At the same time, CPT happens at the center of the magneto-optical resonance ($\Omega=0$) and there is almost no any population of the exited state $F_e=1$ and the essential optical pumping of the nonresonant level $F_g=2$ does not occur – probe-wave absorption at $\Omega=0$ remains relevant. These considerations make it obvious that a closed transition should not demonstrate the ultra-high contrast effect. By the way, physical origin of the new effect is close to that studied in the paper [53], where optical pumping under counterpropagating light waves with parallel linear polarizations led to some increase of magneto-optical EIA amplitude as well, but in not so much level as in the present paper.

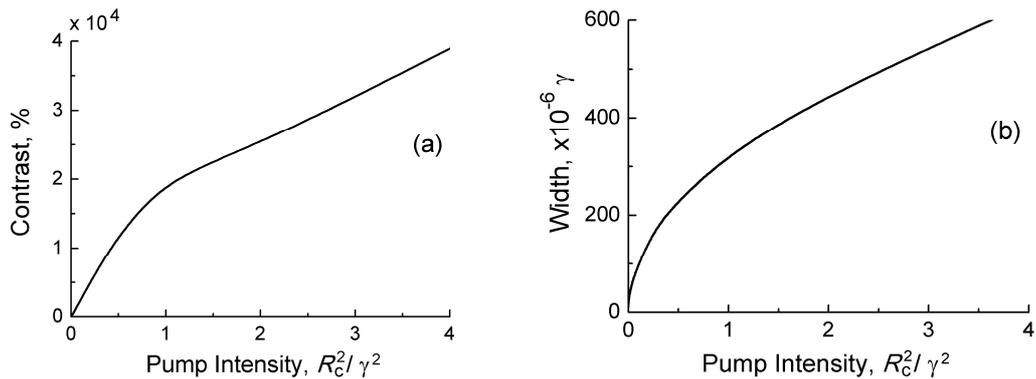

**Figure 4.** Contrast $C_\alpha$ (a) and width (b) of subnatural EIA resonance in their dependence on the pump-wave intensity. Rabi frequency of the probe beam is $R_p=0.01\gamma$.

In a real experiment, as usual, the transmission signal, such as (4), is measured rather than the coefficient $\alpha$. It is due to the coefficient of probe-wave absorption can be hardly extracted directly from the transmitted signal or the total fluorescence of vapour cell. In case of transmitted signal (4) EIA resonance retains narrow width, but it may be far from having a high contrast (see figure 3b). In other words, the amplitude of central nonlinear spike at figure 3b is much less than the background level.

For the resonances in transmission signal the contrast better to be defined as $C_t=(B_1/B_2)\times 100\%$ (see figure 3b). In particular, $C_t\approx 10\%$ for the resonance at figure 3b. Actually, the problem with low EIA contrast in transmission (4) can be easily solved by increasing the product $n\times L$. For instance, we can take a typical value $L=5$ cm (as at figure 3). Then, for example, by increasing the temperature of vapour cell, the atomic density can be brought to the value $n=5\times 10^{11}$ cm$^{-3}$. At these conditions $C_t$ tends to be close to 100% ($B_1\approx B_2$) simultaneously with sufficiently narrow width $\Delta_{EIA}\approx 700$ Hz (see figure 5a).

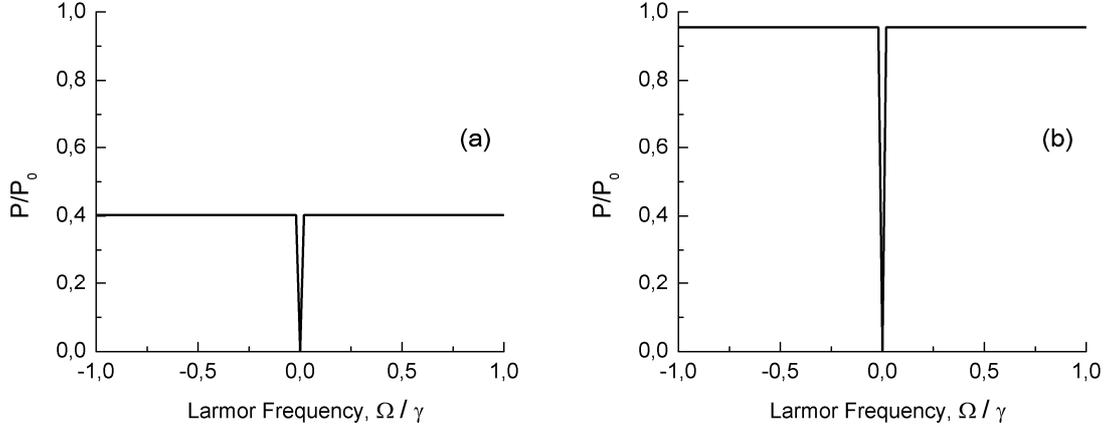

**Figure 5.** (a) High-contrast EIA resonance in the probe-wave transmission signal at increased atomic density $n=5\times 10^{11}$ cm$^{-3}$ and pump-wave Rabi frequency $R_c=0.15\gamma$. Other parameters are as for the figure 3.

In spite of good contrast $C_t\approx 100\%$ at figure 5a, the background level may be not so small. It seems to be interesting to establish a new task: how can we get full probe-beam transmission in vicinity of profile center ($\Omega>\Delta_{EIA}$) and almost absolutely zero transmission at exact resonance? Reduction of the background can be achieved by increasing intensity of the pump field (as more atoms are pumped into the level $F_g=2$ as the background signal is smaller). If we increase pump-wave Rabi frequency just to the value $R_c=0.7\gamma$, it is already enough to greatly reduce the background and obtain 95% of output power (see figure 5b). At that the transmitted power at the center of the resonance ($\Omega=0$) becomes no more than 0.4% from the initial one. Since the Rabi frequency is increased, EIA width is also increased up to $\Delta_{EIA}\approx 3.3$ kHz. For convenience, this regime may be named as "magneto-optical switch", while the regime with $C_t\approx 100\%$, considerable background level and ultra-narrow width ($\Delta_{EIA}<1$ kHz, as at figure 5a) may be denoted just as "high-contrast" one.

In the present paper we have attracted attention only to the case of linearly polarized waves, but the resonance of "bright" type on "dark" atomic transitions can be also observed for elliptically polarized waves [37]. Besides, the special attention should be paid for studying the influence of stray magnetic field and its inhomogeneity along vapour cell for the proposed configuration. Indeed, the spectroscopic signal in our case is formed by scanning the static longitudinal magnetic field. Unfortunately, in the real experiment there is no possibility to absolutely shield an external field. Moreover, the internal solenoid, producing the longitudinal magnetic field, cannot be ideal and arranged absolutely parallel to the beam's propagation direction. All these factors lead to the presence of stray magnetic field with heterogeneous amplitude and direction. The estimates show that the stray field should not exceed the value of 0.5 mG to observe the ultra-narrow EIA resonances ($0.2<\Delta_{EIA}<1$ kHz). This is sufficiently strict requirement, which may be satisfied, for example, with applying special multilayer magnetic shield. Also, the buffer gas pressure should not be too large, when the collisional broadening becomes equal to the hyperfine splitting of the exited states of the transition (about 817 MHz for the D$_1$ line of $^{87}$Rb). This case should be studied separately.

## 3. Summary

To conclude, we would like to note main results. The new method for observing the narrow-width and high-contrast magneto-optical resonances of electromagnetically induced absorption has been proposed. It is based on the configuration of electromagnetic fields composed of two counterpropagating laser beams and static magnetic field applied along the wave vectors. The configuration allows using buffer gas or cell with antirelaxation coating for enhancing in great level the properties of the subnatural-width resonance. It has been suggested to exploit "dark" types of open atomic transitions and perpendicular linear polarizations of the waves to obtain narrow and high-contrast "bright" resonances (EIA). Moreover, openness of a transition can cause the ultra-high contrast of subnatural resonance in the probe-wave absorption coefficient (up to $10^4$%), when its profile contains almost only nonlinear EIA signal without any broader background.

The two principal regimes of probe-wave transmission signal have been analyzed. The first one allows obtaining high contrast (≈100%) and ultra-narrow width (< 1 kHz) of EIA resonances. The second regime may be called as "magneto-optical switch", because at exact resonance condition (zero magnetic field) transmitted signal equals to zero, while near the resonance transmission is almost 100%. In this regime the EIA contrast is also equals to 100%, but the width increases (units of kHz) in comparison with the first regime. In general, width of EIA resonance in the configuration proposed is limited by time of life of anisotropy in the ground state of an atom. Potentially, in the case of buffer gas the width can be as small as several tens of Hz.

It is interesting to note, that the effect of high contrast of observed magneto-optical EIA resonances makes open atomic transitions preferable to closed ones. At the same time, optical pumping towards non-resonant ground hyperfine sublevels for open transitions usually negatively affects on the properties of EIT resonances [46] as well as EIA ones in the regular observing schemes [21,47]. Therefore, openness of a transition gives significant advantage to EIA signals in proposed scheme over EIT ones or EIA in the regular schemes of observation.

In the proposed scheme the signal is formed by scanning the static magnetic field. However, the special interest may be attracted for making the scheme all-optical, i.e. with frequency scanning. In that case high properties of EIA resonance in transmission signal should also accompany very steep dispersion, what could found interesting applications in nonlinear optics and optical communications.

**Acknowledgments**

The work was partially supported by the Russian Foundation for Basic Research (## 14-02-00806, 14-02-00712, 14-02-00939, 12-02-00403, 12-02-00454), Ministry of Education and Science of the Russian Federation, Russian Academy of Sciences and Presidium of the Siberian Branch of RAS, Russian Quantum Center ("Skolkovo", Moscow region). D.V.B. also acknowledges the Presidential Council and Ministry of Education and Science of the Russian Federation for the grant # MK-4680.2014.2.